\documentstyle[preprint,aps]{revtex}
\begin{document}
\draft
\preprint{
\parbox{4cm}{
\baselineskip=12pt
TMUP-HEL-9509\\
UT-731\\
October, 1995\\
\hspace*{1cm}
}}
\title{Possible explanation of $Zbb$ vertex anomaly\\
       in one-family technicolor models}
\author{N.~Kitazawa
 \thanks{e-mail: kitazawa@musashi.phys.metro-u.ac.jp}}
\address{Department of Physics, Tokyo Metropolitan University,
         Hachioji-shi, Tokyo 192-03, Japan
\\
         and\\
         Department of Physics, Yale University,
         PO Box 208120, New Haven, CT 06520, USA}
\author{and}
\author{T.~Yanagida
 \thanks{e-mail: yanagida@danjuro.phys.s.u-tokyo.ac.jp}}
\address{Department of Physics, University of Tokyo,
         Bunkyo-ku, Tokyo 113, Japan}
\maketitle
\begin{abstract}
In technicolor theories
 a large radiative correction to the $Zbb$ vertex generally emerges,
 because of the large top quark mass and $SU(2)_L$ gauge symmetry.
The correction can give us an explanation
 for the deviation of the experimental value of $R_b \equiv \Gamma_b/\Gamma_h$
 from the standard-model prediction.
However, generally the $T$ parameter becomes unacceptably large.
We show that
 in the one-family technicolor model recently proposed by the authors
 all the $S$, $T$, and $U$ parameters
 can be consistent with the experimental bounds
 while the observed anomaly of $R_b$ is naturally explained.
\end{abstract}
\newpage

The measurement of the quantity $R_b \equiv \Gamma_b / \Gamma_h$ at LEP
 shows a large deviation from the prediction of the standard model.
The measured value $R_b = 0.2202 \pm 0.0020$
 deviates at 2-$\sigma$ level from the standard-model prediction
 $R_b = 0.2157$ ($m_t = 175$ GeV)\cite{LEP,HHKM}.
The recent preliminary experimental value $R_b = 0.2219 \pm 0.0017$
 shows the deviation at 3.6-$\sigma$ level\cite{LEP-2}.
This may be a signature of new physics.

It has been pointed out that
 the ``sideways'' gauge boson of extended technicolor (ETC) theories
 generates a significant correction to the $Zbb$ vertex\cite{chivukula}.
The reason is that
 the relatively light sideways gauge boson
 associated with the top quark mass generation
 must couple with the left-handed bottom quark
 according to the $SU(2)_L$ symmetry.
The ``diagonal'' gauge boson which appears in most ETC models
 also generates a correction to the $Zbb$ vertex\cite{kitazawa}.
The magnitude of this diagonal contribution
 is comparable with the sideways contribution
 and the sign is opposite\cite{wu}.
The sideways and the standard-model contributions make $R_b$ small,
 while the diagonal contribution makes it large.
Therefore,
 if the diagonal contribution is large enough
 to cancel out the other two contributions,
 the LEP result can be explained.
It has been shown that
 the diagonal contribution can naturally explain the LEP result of $R_b$
 in some ETC models\cite{hagiwara-kitazawa}.

However, the contribution of the diagonal ETC boson
 to the $T$ parameter\cite{peskin-takeuchi}
 generally becomes large in comparison with the experimental bound,
 because of the large weak isospin violation in the couplings
 between the diagonal ETC boson and the right-handed fermions
 (top and bottom quarks and techni-fermions).
It has been shown that
 it is almost impossible in the naive ETC theory
 to explain the LEP result of $R_b$
 without conflict with the experimental bound
 on the $T$ parameter\cite{yoshikawa}.
Some mechanisms which generate a negative contribution to the $T$ parameter
 are needed to explain the experimental value of $R_b$.

We stress here the fact that
 the large contributions
 to both the $Zbb$ vertex correction and the $T$ parameter
 are not peculiar to a specific ETC theory
 but generally result in any technicolor scenario
 with some underlying physics at TeV scale for the fermion mass generation.
To generate the heavy top quark,
 at least the effective four-fermion interaction
\begin{equation}
 {\cal L}^{t_R} = {1 \over {\Lambda^2}}
                  \Big( {\bar Q}_L U_R \Big)
                  \Big( {\bar t}_R \psi_L \Big) + \mbox{h.c.}
\label{op-top-mass}
\end{equation}
 must be emerged,
 where $Q_L = ( U_L \ D_L )^T$ and $\psi_L = ( t_L \ b_L )^T$
 are the $SU(2)_L$ doublets, and $U$ and $D$ are the techni-quarks.
If we consider the QCD-like one-family technicolor model,
 the scale $\Lambda \simeq 260$GeV with $m_t=175$GeV.
Here,
 we used the relation
 $\langle {\bar U} U \rangle \simeq 4 \pi F_\pi^3$
 (taking the number of the technicolor $N_{TC}=3$)\cite{manohar-georgi},
 and set $F_\pi \simeq 125$GeV
 ($F_\pi$ is the decay constant of the composite Nambu-Goldstone boson).
Namely,
 some unknown dynamics has to emerge
 at the scale of order $\Lambda \simeq 260$GeV.

In general,
 other kind of effective four-fermion interactions
 also emerge at the scale $\Lambda$.
The effective four-fermion interactions with $U_R$
\begin{eqnarray}
 {\cal L}^{U_R}_1 &=& {{g^U_1} \over {\Lambda^2}}
                      \Big( {\bar \psi}_L U_R \Big)
                      \Big( {\bar U}_R \psi_L \Big),
\label{op-Zbb-top}
\\
 {\cal L}^{U_R}_2 &=& {{g^U_2} \over {\Lambda^2}}
                      \Big( {\bar U}_R \gamma^\mu U_R \Big)
                      \Big( {\bar U}_R \gamma_\mu U_R \Big),
\label{op-T-top}
\end{eqnarray}
 are expected to emerge with the scale $\Lambda$ and $g^U_{1,2} \sim 1$.
The interaction of eq.(\ref{op-Zbb-top})
 gives the correction to the $Zb_Lb_L$ vertex,
 and the interaction of eq.(\ref{op-T-top})
 gives the contribution to the $T$ parameter.

The order of the correction to the $Zb_Lb_L$ coupling $g_L^b$
 (at tree-level in the standard model
  $g_L^b = g_Z \{ -{1 \over 2} + {1 \over 3} s^2 \}$,
  where $g_Z = \sqrt{4 \pi \alpha / c^2 s^2}$,
  and $s$ and $c$ denote the sine and cosine
  of the Weinberg angle, respectively)
 due to the interaction of eq.(\ref{op-Zbb-top}) is estimated as
\begin{equation}
 \left| (\delta g_L^b)_{\Lambda} \right|
  \simeq {{g_Z} \over 8}{{F_\pi^2} \over {\Lambda^2}} |g^U_1|
  \simeq 0.021 |g^U_1|,
\label{general-zbb}
\end{equation}
 which should be compared with the experimental bound
 on the {\it total} correction to the coupling $g_L^b$
 from $R_b$ data alone\cite{matsumoto}
\begin{equation}
 \delta g_L^b = - 0.0004 \pm 0.0019
\label{experiment-glb}
\end{equation}
 and the standard-model contribution
 $(\delta g_L^b)_{SM} = 0.0037$\cite{HHKM}.
If the magnitude of the coupling $g^U_1$
 is of the order of $10^{-1}$ and the sign is negative,
 the contribution of the standard model can be canceled out
 and the experimental value of eq.(\ref{experiment-glb}) can be explained.
In this estimation we used the method
 in which the techni-fermion current is replaced by the corresponding current
 in the low energy effective theory\cite{chivukula}.
The significantly large correction of eq.(\ref{general-zbb}) comes from
 the large top quark mass and $SU(2)_L$ symmetry.

The order of the contribution to the $T$ parameter
 due to the interaction of eq.(\ref{op-T-top}) is estimated as
\begin{equation}
 \left| T_{\Lambda} \right|
  \simeq {{g_Z^2 N_C^2} \over {16 \alpha}}
         {{F_\pi^4} \over {m_Z^2 \Lambda^2}} |g^U_2|
  \simeq 17 |g^U_2|,
\end{equation}
 where $N_C=3$.
This value should be compared with the experimental bound\cite{matsumoto}
\begin{equation}
 T = 0.34 \pm 0.20
\label{experiment-T}
\end{equation}
 for $m_t=175$GeV and $m_H=1$TeV.
We used also the method of the current replacement
 with the factorization hypothesis\cite{yoshikawa}.
This extremely large contribution to the $T$ parameter comes from
 the large weak isospin violation
 in the mechanism of the fermion mass generation.
Even if
 $|g^U_2|$ is of the order of $10^{-1}$ as we expect for $|g^U_1|$,
 the magnitude of $|T_\Lambda|$ is still much larger
 than the experimental bound.

Therefore,
 having large contributions
 to the $Zb_Lb_L$ vertex correction and the $T$ parameter
 is the generic in the technicolor scenario,
 if there are no special cancelations or suppressions.
If the large deviation of the experimental value of $R_b$
 from the the standard-model prediction
 is naturally explained in the technicolor scenario
 with the appropriate value of $g^U_1$,
 some mechanisms to generate the negative contribution to the $T$ parameter
 and/or some mechanisms which suppress the magnitude of $g^U_2$
 are needed.

In the one-family technicolor model
 recently proposed by the authors\cite{kitazawa-yanagida}
 we have a mechanism for generating large negative $T$ parameter contributions
 while both the $S$ and $U$ parameters
 are consistent with the experimental bounds.
The one-family technicolor model
 has the separate structure of the technicolor gauge interactions
 $SU(3)^Q_{TC} \times SO(3)^L_{TC} \times U(1)^{TF}_{B-L}$,
 where $SU(3)^Q_{TC}$ couples only with the techni-quarks
 in the triplet representation,
 $SO(3)^L_{TC}$ couples only with the techni-leptons
 in the triplet representation,
 and $U(1)^{TF}_{B-L}$ denotes the techni-$(B-L)$ symmetry
 which is spontaneously broken.
This structure realizes the minimum value of the number of technicolors
 $N_{TC}$ (the smaller $N_{TC}$ means the smaller $S$ parameter)
 while the techni-leptons belong to the real representation
 of the technicolor gauge group,
 which is needed to have the gauge invariant Majorana mass
 of the right-handed techni-neutrino.
The Majorana mass
 can generate the negative contribution to the $T$ parameter.

In the following
 we estimate how large the correction to $g_L^b$ may be
 keeping the values of the $S$, $T$, and $U$ parameters
 consistent with the experimental bounds
 in this one-family technicolor model.
We consider the effective model of the fermion mass generation
 in ref.\cite{hagiwara-kitazawa},
 since it can naturally explain the observed anomaly of $R_b$.
Although the effective model is inspired by an ETC theory,
 it is not itself an ETC model.
We do not always restrict
 the fermion mass generation mechanism to the ETC theory,
 but imagine some underlying physics at TeV scale
 for the fermion mass generation.

We obtain the effective model of the fermion mass generation
 in the following way.
First we consider the naive ETC scenario with no weak isospin violation,
 for example,
 $SU(4)^Q_{ETC} \times SU(4)^L_{ETC} \times U(1)^{ETC}_{B-L}
  \rightarrow
  SU(3)^Q_{ETC} \times SO(3)^L_{ETC} (\times U(1)^{TF}_{B-L})_{\rm broken}$.
Then we replace
 the gauge coupling of the massive sideways and diagonal ETC bosons
 by the effective couplings
 so that the weak isospin violation is included.
Therefore,
 this effective model is not a gauge theory,
 but we assume that the massive particles
 like the sideways and diagonal ETC bosons in this model
 emerge due to the underlying physics at the TeV scale.
In other words,
 we assume that a part of the low-energy effective interactions
 of the underlying theory
 is well described by the following effective model.

The exchanges of some massive particles
 like the sideways ETC gauge boson
 give the effective four-fermion interactions
\begin{eqnarray}
 {\cal L}_1^S &=& - {{g_t^2} \over {M_S^2}}
                    \Big( {\bar Q}_L \gamma^\mu \psi_L \Big)
                    \Big( {\bar t}_R \gamma_\mu U_R \Big) + \mbox{h.c.},
\label{4F-sideways-mt}
\\
 {\cal L}_2^S &=& - {{(g_t \xi_t)^2} \over {M_S^2}}
                    \Big( {\bar Q}_L \gamma^\mu \psi_L \Big)
                    \Big( {\bar \psi}_L \gamma_\mu Q_L \Big),
\label{4F-sideways-zbb}
\end{eqnarray}
 where $M_S$ denotes the typical mass of the exchanged particles,
 and $g_t \xi_t$ and $g_t / \xi_t$ are the effective coupling constants
 of the left-handed current and the right-handed current with $t_R$,
 respectively.
We assume that the perturbative condition
\begin{equation}
 {{(g_t \xi_t)^2} \over {4\pi}} < 1
\quad \mbox{and} \quad
 {{(g_t / \xi_t)^2} \over {4\pi}} < 1
\label{perturbative-condition}
\end{equation}
should be satisfied.
The possible range of $\xi_t^2$ is determined by this condition,
 when the value of $g_t^2$ is given.
The mass of the top quark
 is generated through the interaction of eq.(\ref{4F-sideways-mt}) as
\begin{equation}
 m_t \simeq {{g_t^2} \over {M_S^2}}
            4 \pi F_\pi^3 \sqrt{{{N_C} \over {N_{TC}}}},
\label{top-mass}
\end{equation}
 where the relation
 $\langle \bar{U} U \rangle \simeq 4 \pi F_\pi^3 \sqrt{N_C / N_{TC}}$
 is used
 (from the naive dimensional analysis\cite{manohar-georgi}
  and the leading $1/N$ behavior).
We can determine the value of $g_t^2$,
 when the value of $M_S$ is given ($N_{TC}=3$ in our model).
The exchanges of some massive particles
 like the diagonal ETC gauge boson give
\begin{equation}
 {\cal L}^D = - {1 \over {M_D^2}} J_D^\mu J_{D\mu},
\label{4F-diagonal}
\end{equation}
 where
\begin{eqnarray}
 J_{D\mu} &=& g_t \xi_t \sqrt{{N_{TC}} \over {N_{TC}+1}}
               \left( a \ {\bar \psi}_L \gamma_\mu \psi_L
                    - b \ {1 \over {N_{TC}}} {\bar Q}_L \gamma_\mu Q_L
               \right)
\nonumber\\
          &+& {{g_t} \over {\xi_t}} \sqrt{{N_{TC}} \over {N_{TC}+1}}
               \left( a \ {\bar t}_R \gamma_\mu t_R
                    - b \ {1 \over {N_{TC}}} {\bar U}_R \gamma_\mu U_R
               \right)
\nonumber\\
          &+& {{g_t} \over {\xi_b}} \sqrt{{N_{TC}} \over {N_{TC}+1}}
               \left( a \ {\bar b}_R \gamma_\mu b_R
                    - b \ {1 \over {N_{TC}}} {\bar D}_R \gamma_\mu D_R
               \right),
\end{eqnarray}
 $M_D$ denotes the typical mass of the exchanged particle,
 and $g_t / \xi_b$ is the effective coupling constant
 of the right-handed current with $b_R$.
The constants $a$ and $b$ equal unity and $M_D = M_S$
 in the naive ETC model based on the one-family technicolor theory.
In general,
 the constants $a$ and $b$ are expected to be of the order of unity
 and $M_D \simeq M_S$.

The interactions of eq.(\ref{4F-sideways-zbb}) and (\ref{4F-diagonal})
 generate the correction to $g_L^b$ (see ref.\cite{hagiwara-kitazawa}):
\begin{eqnarray}
  (\delta g_L^b)_{\Lambda} &=&
 \left( \xi_t^2 - a b {{2N_C} \over {N_{TC}+1}} \right)
 {{m_t} \over {16 \pi F_\pi}} \sqrt{{{N_{TC}} \over {N_C}}} g_Z
\nonumber\\
 &\simeq& \left( \xi_t^2 - {3 \over 2} a b  \right) \cdot 2.1 \times 10^{-2},
\end{eqnarray}
 where we have assumed that $M_D \simeq M_S$ and used eq.(\ref{top-mass}).
The first term in the brackets
 comes from the interaction of eq.(\ref{4F-sideways-zbb}),
 and the second term comes from the interaction of eq.(\ref{4F-diagonal}).
The interactions of eq.(\ref{4F-diagonal})
 generates large contribution to the $T$ parameter
 (see ref.\cite{kitazawa-2}):
\begin{equation}
 T_{\Lambda} = {1 \over {16 c^2 s^2}}{{m_t F_\pi} \over {m_Z^2}}
            {{N_C+1} \over {N_{TC}+1}} \sqrt{{N_C} \over {N_{TC}}}
            \left( 1 - {{m_b} \over {m_t}} \right)^2
            {{b^2} \over {\xi_t^2}}
         \simeq 0.93 \cdot {{b^2} \over {\xi_t^2}}.
\end{equation}
(In most ETC models
  the massive ETC gauge boson
  which belongs to the adjoint representation of the technicolor gauge group
  generates unacceptably large positive contribution
  to the $T$ parameter.
 However,
  we assume that our underlying theory
  does not have such a dangerous adjoint vector bosons\cite{evans}.)

In the one-family technicolor model of ref.\cite{kitazawa-yanagida}
 the value of the $T$ parameter depends on the values
 of the Dirac masses of the techni-neutrino and techni-electron
 $m_N$ and $m_E$, respectively.
The values of $m_N$ and $m_E$ are dynamically determined,
 if we fixed the values of the Majorana mass
 of the right-handed techni-neutrino $M$,
 the mass and coupling of the $U(1)^{TF}_{B-L}$ gauge boson
 ($m_{B-L}$ and $\alpha_{B-L}$, respectively),
 and the coupling of the $SO(3)^L_{TC}$.
The calculation of the vacuum energy
 in the one gauge boson exchange approximation shows
 that the difference between the values of $m_N$ and $m_E$ is about $60$GeV,
 when $M = 250$GeV, $m_{B-L}=250$GeV, and $\alpha_{B-L}=0.3$.
These values of $M$, $m_{B-L}$, and $\alpha_{B-L}$,
 and the value $\omega=0.07$ which is the strength of the kinetic mixing
 between the $U(1)^{TF}_{B-L}$ and $U(1)_Y$ gauge bosons are selected
 so that the correct electroweak symmetry breaking occurs
 and the values of the $S$ and $U$ parameters are consistent
 with the experimental bounds.
Both the Majorana mass $M$
 and the tree-level kinetic mixing $\omega$
 play an important role to get the large negative contribution
 to the $S$ parameter through the $U(1)^{TF}_{B-L}$ gauge boson exchange.
Large kinetic mixing between the $W^3$ and $U(1)_Y$ gauge bosons
 comes from the loop-level mass mixing
 between the $W^3$ and $U(1)^{TF}_{B-L}$ gauge bosons
 due to the Majorana mass,
 and the tree-level kinetic mixing $\omega$
 between the $U(1)^{TF}_{B-L}$ and $U(1)_Y$ gauge bosons.

If we take $m_N=340$GeV and $m_E=400$GeV,
 we have $S_{TC} \simeq -0.0092$, $T_{TC} \simeq -0.21$,
 and $U_{TC} \simeq 0.022$.
These values of $S_{TC}$ and $U_{TC}$
 are consistent with the experimental bounds of
 $S = 0.068 \pm 0.20$ and $U = -0.41 \pm 0.50$
 with $m_t=175$GeV and $m_H=1$TeV\cite{matsumoto}.
Although a smaller difference of $m_N$ and $m_E$
 gives a smaller value of the $T$ parameter,
 $m_E-m_N < 60$GeV is rather hard to reconcile with $M = 250$GeV.

Now we estimate
 how large the value of $- (\delta g_L^b)_{\Lambda}$ can be
 in this one-family technicolor scenario.
A Negative contribution to $g_L^b$
 corresponds to a positive contribution to $R_b$
 which is suggested from the experiment.
By using the experimental bound of eq.(\ref{experiment-T}),
 we can obtain the upper bound on the parameter $b^2/\xi_t^2$ as
\begin{equation}
 T_{\Lambda} + T_{TC}
  = {1 \over {16 c^2 s^2}}{{m_t F_\pi} \over {m_Z^2}}
    {{N_C+1} \over {N_{TC}+1}} \sqrt{{N_C} \over {N_{TC}}}
    \left( 1 - {{m_b} \over {m_t}} \right)^2
    {{b^2} \over {\xi_t^2}} - 0.21
  < 0.34 + 0.20.
\end{equation}
We have $b^2/\xi_t^2 < 0.80$.
{}From this bound we obtain the lower bound on the $(\delta g_L^b)_{\Lambda}$:
\begin{equation}
  (\delta g_L^b)_{\Lambda} >
 \left( {{b^2} \over {0.80}} - a b {{2N_C} \over {N_{TC}+1}} \right)
 {{m_t} \over {16 \pi F_\pi}} \sqrt{{{N_{TC}} \over {N_C}}} g_Z
 \simeq b ({b \over {0.80}} - {3 \over 2} a) \cdot 0.021.
\end{equation}
In the case of $a=b=1$,
 we have $(\delta b_L^b)_{\Lambda} > - 0.0053$,
 or $(\delta R_b)_{\Lambda} < 0.0069$.
The perturbative condition eq.(\ref{perturbative-condition})
 is satisfied in this case with $M_S=1$TeV.
Then we obtain the bound on the {\it total} correction as
 $\delta g_L^b = (\delta g_L^b)_{SM} + (\delta g_L^b)_{\Lambda} > -0.0016$.
Therefore the experimental value of eq.(\ref{experiment-glb})
 can be explained in this model
 keeping the $S$, $T$, and $U$ parameters
 consistent with the experimental bounds.
Both the Majorana mass $M$ and the tree-level kinetic mixing $\omega$
 play the important role of generating the large negative contribution
 to the $S$ parameter
 through the $U(1)^{TF}_{B-L}$ gauge boson exchange.
The Majorana mass $M$ is needed to have the mass mixing
 between the $U(1)^{TF}_{B-L}$ gauge boson and $W^3$,
 and the tree-level mixing $\omega$ makes this contribution large.
If there is no negative contribution to the $T$ parameter, namely $T_{TC}=0$,
 the experimental value of eq.(\ref{experiment-glb}) can not be explained,
 because we have the bounds $(\delta g_L^b)_{\Lambda} > 0.0047$
 and $\delta g_L^b > 0.0084$ in this case.
As long as the model dependent parameters $a$ and $b$
 satisfy the condition $b ({b \over {0.80}} - {3 \over 2} a) < 0$,
 we obtain the negative contribution to $g_L^b$
 (the positive contribution to $R_b$)
 which is favored by the experiment.

N.K. is grateful to Prof. A.Chodos
 for the hospitality during my staying at Yale university,
 and also wish to thank to N.Evans, S.D.H.Hsu, and S.B.Selipsky
 for the stimulating discussions.


\begin{references}
\bibitem{bottom}
Similar effective four-fermion interactions
  like eqs.(\ref{op-Zbb-top}) and (\ref{op-T-top}) with $D_R$
  are also expected to emerge.
 But they do not give the dominant effect
  on the $Zbb$ vertex and the $T$ parameter,
  since the scale of the interactions are expected
  to be larger than $\Lambda$.
\end{references}
\end{document}